%% file: main.tex
\documentclass[sigconf,natbib=true]{acmart}

\usepackage{booktabs}
\usepackage{enumerate}
\usepackage{enumitem}

\usepackage[linesnumbered,ruled,vlined]{algorithm2e}
\usepackage{algpseudocode}

\usepackage{times}
\usepackage{soul}
\usepackage{url}
\usepackage{tabularx}
\usepackage{multirow}
\usepackage{eucal}
\usepackage{amsfonts}
\usepackage{subfig}
\usepackage{titlesec}
\usepackage{flushend}
\usepackage{bm}
\usepackage{amsthm,amsmath,amssymb}
\usepackage{mathrsfs}
\usepackage{verbatim}
\usepackage{xcolor}
\usepackage{pifont}
\makeatletter
\newcommand\figcaption{\def\@captype{figure}\caption}
\newcommand\tabcaption{\def\@captype{table}\caption}
\makeatother

\newcommand\modelname{{I-Pro}} 

\definecolor{blue0}{RGB}{3,60,163}
\definecolor{green0}{RGB}{83,153,55}

\AtBeginDocument{%
  \providecommand\BibTeX{{%
    \normalfont B\kern-0.5em{\scshape i\kern-0.25em b}\kern-0.8em\TeX}}}

\setcopyright{acmcopyright}
\copyrightyear{2022}
\acmYear{2022}
\acmDOI{XXXXXXX.XXXXXXX}

\acmConference[SIGIR'22]{Make sure to enter the correct
  conference title from your rights confirmation emai}{July 11–15, 2022}{Madrid, Spain}
\acmBooktitle{SIGIR'22: July 11–15, 2022, Madrid, Spain} 
\acmPrice{15.00}
\acmISBN{978-1-4503-XXXX-X/18/06}

\begin{document}


\title[A New Paradigm for Proactive Dialogue Policy]{Interacting with Non-Cooperative User:\\ A New Paradigm for Proactive Dialogue Policy}

\author{Wenqiang Lei$^{*}$$^1$,~~Yao Zhang$^*$$^2$,~~Feifan Song$^3$,~~Hongru Liang$^1$,~~Jiaxin Mao$^4$\\ Jiancheng Lv$^1$,~~Zhenglu Yang$^2$,~~Tat-Seng Chua$^5$ }

\affiliation{%
  \institution{$^1$College of Computer Science, Sichuan University, China~~~$^2$TKLNDST, CS, Nankai University, China}
  \institution{$^3$MOE Key Lab of Computational Linguistics, Peking University, China}
  \institution{$^4$GSAI, Renmin University of China, China~~~$^5$National University of Singapore, Singapore}
  \city{ }
  \country{ }}
\email{ wenqianglei@gmail.com,   yaozhang@mail.nankai.edu.cn,   songff@stu.pku.edu.cn,   lianghongru@scu.edu.cn}
\email{maojiaxin@ruc.edu.cn,   lvjiancheng@scu.edu.cn,   yangzl@nankai.edu.cn,   chuats@comp.nus.edu.sg}

\thanks{$^*$Both authors contributed equally to this research.}
\thanks{Wenqiang Lei is the Corresponding Author.}

\renewcommand{\shortauthors}{Wenqiang Lei and Yao Zhang, et al.}

\begin{abstract}
Proactive dialogue system is able to lead the conversation to a goal topic and has advantaged potential in bargain, persuasion and negotiation. 
Current corpus-based learning manner limits its practical application in real-world scenarios. 
To this end, we contribute to advance the study of the proactive dialogue policy to a more natural and challenging setting, i.e., interacting dynamically with users.
Further, we call attention to the non-cooperative user behavior --- the user talks about off-path topics when he/she is not satisfied with the previous topics introduced by the agent. 
We argue that the targets of reaching the goal topic quickly and maintaining a high user satisfaction are not always converge, because the topics close to the goal and the topics user preferred may not be the same. Towards this issue, we propose a new solution named \textbf{\modelname} that can learn \underline{Pro}active policy in the \underline{I}nteractive setting. Specifically, we learn the trade-off via a learned goal weight, which consists of four factors~(dialogue turn, goal completion difficulty, user satisfaction estimation, and cooperative degree). 
The experimental results demonstrate \modelname{} significantly outperforms baselines in terms of effectiveness and interpretability.

\end{abstract}




\maketitle

\input{body/intro}

\input{body/framework}

\input{body/model}

\input{body/experiment}

\input{body/relatedwork}

\input{body/conclusion}

\section*{Acknowledgments}
This work was supported in part by the scholarship from China Scholarship Council (CSC) under Grant No.202006200134; in part by the China Postdoctoral Science Foundation under Grant No.2021TQ0222 and No.2021M700094; in part by the Fundamental Research Funds for the Central Universities of China; and in part by the Beijing Academy of Artificial Intelligence (BAAI).


\bibliographystyle{ACM-Reference-Format}
\bibliography{sample-base,reference}


\end{document}

%% file: body/intro.tex
\section{Introduction}
\label{intro}

\begin{figure*}[t]
  \centering
  \includegraphics[width=0.99\textwidth]{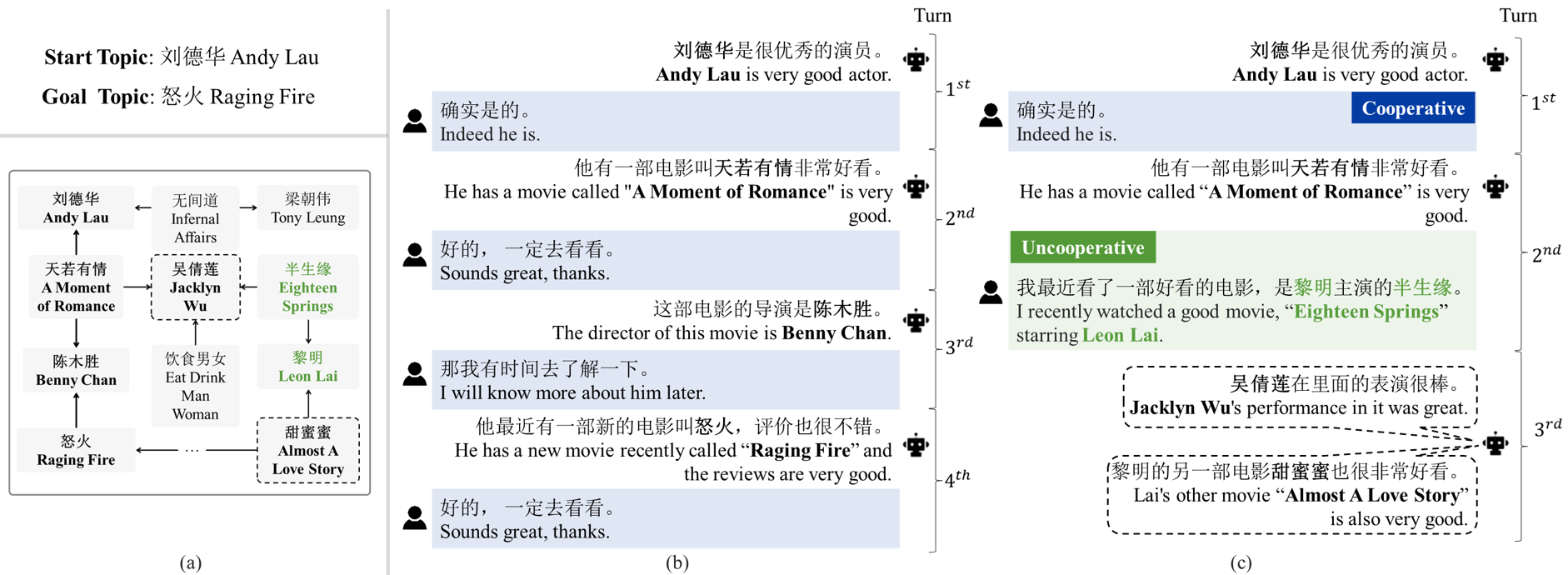}
  \caption{
    Illustration of proactive dialogue: the agent targets at leading the conversation from the start topic ``Andy Lau'' to the goal topic ``Raging Fire'', based on (a) a movie-related KG fragment.
    (b) A successful conversation, in which the agent leads the conversation to the goal topic. This scenario is ideal because the user always behaves {\color{blue0}cooperatively}.
    (c) A realistic conversation fragment. The {\color{green0}non-cooperative} user behavior takes the conversation out of the agent’s control, so the agent needs to reconsider which topic (in the dotted box) to introduce to the user at the $3^{rd}$ turn.
    Since the KG used in this work is in Chinese, both Chinese and its corresponding English translation are shown here.
  Best viewed in color.
  }
  \label{fig:intro1} 
  \end{figure*} 

 Proactive dialogue agent aims to lead the conversation with a user from the start topic~(``Andy Lau'') to the goal topic~(``Raging Fire'') through chatting with the user~\cite{2019Proactive}, as shown in Figure~\ref{fig:intro1}.  
 This task has great potential in scenarios like bargain~\cite{he-etal-2018-decoupling}, persuasion~\cite{wang-etal-2019-persuasion,dutt2021resper} and negotiation~\cite{lewis-etal-2017-deal,yang2020generating}. Current solutions~\cite{2019Proactive,Hao2020Multi,bai2021learning,wuzhou2021augmenting,zhu2021proactive} follow the \textbf{corpus-based learning setting} --- given a knowledge graph~(KG), a goal topic and a dialogue context between two human~(e.g., the leader and the follower in the DuConv corpus~\cite{2019Proactive}), the agent is required to predict a topic of the next turn and generate a response based on this topic. However, turn-level policy might not align to the conversation-level policy well~\cite{NEURIPS2019_fc981212,zhang-etal-2021-dynaeval}. Thus, we argue that the corpus-based learning setting is insufficient to meet the ultimate end that the agent is capable to chat with the user dynamically. In this work, we take one step further to scrutinize proactive dialogue policy in the \textbf{interactive setting}. 
 This is a more natural but challenging setting where the agent is required to optimize long-term goal during the dynamic interaction.




Intuitively, the most straightforward solution for the policy would be a shortest path from the start to the goal, as presented in Figure ~\ref{fig:intro1}~(b).
However, the users may non-cooperatively introduce off-path topics. It usually happens when they are not satisfied with the current topic
~\cite{su2018user}. Take Figure ~\ref{fig:intro1} (c), as an example, when the user is unsatisfied with the previous topics introduced by the agent, he/she talks about off-path topics~(``Eighteen Springs'' and ``Leon Lai'') in the 2$^{nd}$ turn. Such non-cooperative user behavior has been seldom studied in prior efforts, but is crucial in the dynamic interaction. First, it can make the conversation out of the agent's control --- 
the agent is unable to introduce the next topic~(``Benny Chan'') in the currently planned shortest path and needs to find a new path to the goal topic. Second, the low satisfaction may hurt the user's engagedness and may even lead the user to terminate the interaction session. This motivates us to pay attention to the non-cooperative user behavior and carefully manage user satisfaction during the conversation.



In this work, we aim to learn proactive dialogue policy in the interactive setting where we call attention to the non-cooperative user behavior. To advance this, we believe the agent should accomplish two targets --- \uppercase\expandafter{\romannumeral1}) \textit{reaching the goal topic quickly} to make the conversation short and \uppercase\expandafter{\romannumeral2}) \textit{maintaining a high user satisfaction} to keep the user engaged during the conversation~\cite{2019Proactive}.
However, these targets cannot always converge. For example, in the $3^{rd}$ turn of Figure~\ref{fig:intro1}~(c), ``Jacklyn Wu'' is closer to the goal but ``Almost A Love Story'' may be more preferred by the user.
Hence, the agent's choice needs to take into account long-term issues --- 
introducing the topic close to the goal is risky to trigger non-cooperative user behaviors, or, conversely, introducing the topic user preferred is likely to make the agent take longer turns to reach the goal. A good proactive agent is expected to achieve the balance between these targets.
Inspired by the aforementioned analyses, 
we build \textbf{\modelname}, a novel proactive agent that can learn \underline{Pro}active policy in the \underline{I}nteractive setting. Specifically,
it employs the deep Q-learning algorithm~\cite{mnih2015human} to train the dialogue policy function, optimizing the reward of faster goal arrival and higher user satisfaction. Towards the desired trade-off, we design a learned goal weight. It is derived from dialogue turn, goal completion difficulty, user satisfaction estimation, and cooperative degree. These four factors comprehensively decide which topic to be introduced in the next turn.
In the experiment, we recur to user simulators to automatically interact with the agent because involving real users is laborious.
Mimicking the user behavior, the simulator decides whether to be cooperative based on the quantitative satisfaction or to choose new topics based on their preference.
We further give the simulator a personality character ``tolerance'' describing the variation of non-cooperative user behaviors. A high-tolerance simulator is less likely to behave non-cooperatively, while a low-tolerance simulator
does the opposite. 
The experiments demonstrate the effectiveness and interpretability of the proposed solution over baselines.
In addition, we observe two interesting insights.
First, the agent distinctly tends to prioritize the target ``reaching the goal topic quickly'' when the number of dialogue turns is large. 
Second, users with different tolerance levels require different dialogue policies. In the face of low-tolerance users, the agent prefers to improve user satisfaction to keep he/she more engaged. 

In summary, the main contributions of this work are as follows:

\begin{itemize}
  [leftmargin=*]

  \item We take the first step to scrutinize proactive dialogue policy in the natural interaction setting, where we call attention to cope with non-cooperative user behavior. 
  
  \item We propose a novel model \textbf{\modelname} that enables a trade-off between the targets of reaching the goal topic quickly and maintaining a high user satisfaction. 

  \item   Our grounded work can serve as a preliminary baseline, and the insightful analysis provides the potential of benefiting further research.

\end{itemize}


%% file: body/framework.tex
\section{Preliminary}
\label{framework}

To investigate the proposed paradigm, we formalize a simple and practical interactive setting.
We hope that this setting will serve as a good foundation for the current exploration and inspire more complex and realistic settings in the future.


Our setting refers to the scenario proposed by \cite{2019Proactive} where agent and user engage in a conversation based on topics in KG. 
Starting from a given topic, based on the KG, the agent introduces new topics to the user one by one, and then the user responses to the topic introduced by the agent in the current turn.
The conversation ends when the goal topic is reached.
Specifically, at the current turn, the agent will introduce a new topic adjacent to the one previously talked about to the user and the user will give comments on the topic.
If the user is satisfied with the current conversation, he/she will behave cooperatively, giving positive comments on the topic introduced by the agent in the immediate previous turn, e.g., \textit{sounds great}.
However, if the user is not satisfied, then he/she will behave non-cooperatively, proactively mentioning new topics that the user prefers to increase his/her satisfaction~\cite{su2018user}, e.g., \textit{I recently watched a good movie, ``Eighteen Springs'' starring Leon Lai}.

\begin{figure}[t]
    \centering
    \includegraphics[width=0.43\textwidth]{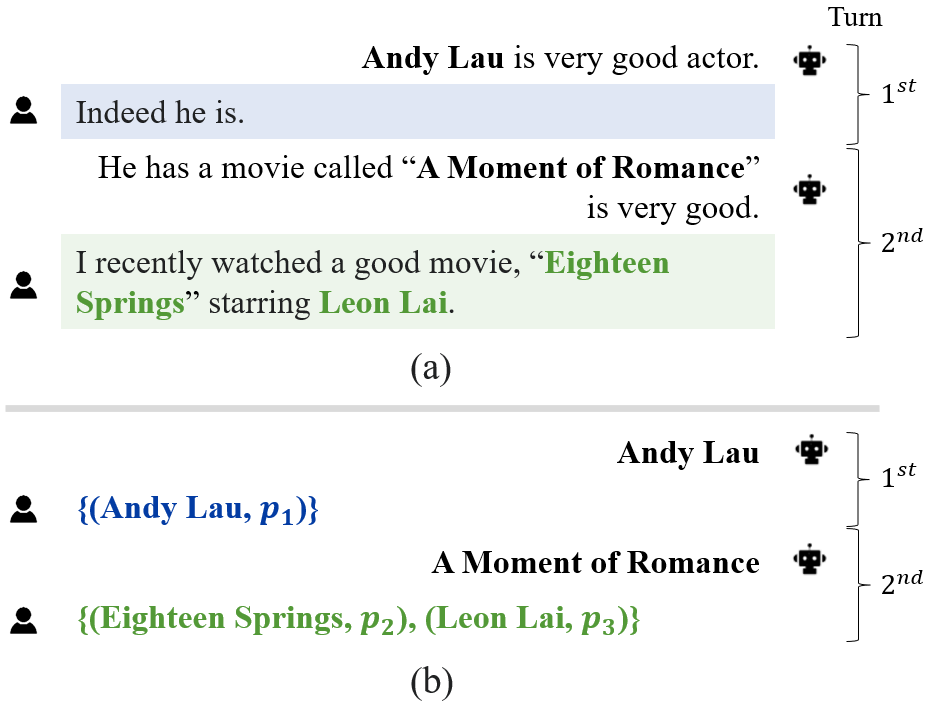}
    \caption{
        The paradigm for proactive dialogue policy: we reuse the $2^{nd}$ turn dialogue in Figure~\ref{fig:intro1} (c) as an example.
        The dialogue is abstracted from natural language level~(a) to topic level~(b). 
        At each turn, the agent introduces one topic to the user. Then the user responds with a list of topics and their corresponding preferences.
    }
    \label{fig:framework} 
    \end{figure}

As this paper focuses on the dialogue policy for the goal leading and user satisfaction, we abstract the whole dialogue process as a topic-level interaction to simplify the problem, as shown in Figure~\ref{fig:framework}. 
Specifically, we assume a third-party natural language understanding technique
 can perfectly detect the topics and user's preference score corresponding to this topic~\footnote{The details of extracting the topic and preference values are not discussed in this paper as it is not the focus of this paper.}.
For example, in the $1^{st}$ turn of the dialogue in Figure~\ref{fig:framework}, the cooperative comments \textit{indeed he is} in response to the topic ``Andy Lau'' can be extracted as $\left \{\right.$(Andy Lau, $p_1$)$\left.\right\}$. 
In the $2^{nd}$ turn, the user utterance \textit{I recently watched a good movie, ``Eighteen Springs'' starring Leon Lai} can be extracted as $\left \{\right.$(Eighteen Springs, $p_2$), (Leon Lai, $p_3$)$\left.\right\}$. $p_1,p_2,p_3$ are preference scores corresponds to ``Andy Lau'', ``Eighteen Springs'', and ``Leon Lai'', respectively.
 In a similar way, we also assume a natural language generation component can perfectly generate an utterance given a chosen topic.
For example, the topic ``A Moment of Romance'' can be generated as \textit{He has a movie called “A Moment of Romance” is very good}. 
Therefore, the conversational policy only needs to focus on topic choosing. 
This perfectly isolates the dialogue policy from natural language understanding and generation, making the conversational agent is dedicated to policy learning.

Formally, the agent starts the conversation with topic $e_{s}$, and leads the conversation to its goal topic $e_{g}$. 
The background KG $\mathcal{G} =\left \{ (e,r,e^\prime ) | e,e^\prime \in \mathcal{E} ,r \in \mathcal{R} \right \} $ is defined as a set of triples with a topic (entity) set $\mathcal{E}$ and relation set $\mathcal{R}$.
Given $T$ dialogue turns, the dialogue history is defined as $H_{T}=\left \{(e_{1}^{a},u_{1}),(e_{2}^{a},u_{2}),...,(e_{T}^{a},u_{T})\right \}$, where $e_{i}^{a}$ is the topic chosen by the agent and $u_{i}$ refers to the user utterance at the $i^{th}$ turn. 
At the $t^{th}$ turn, given the goal topic $e_{g}$ and the dialogue history $H_{t-1}$, the agent chooses a new topic $e_{t}^{a}$ from the candidate topic set~${C}_t$.
Following prior works~\cite{2017He,moon-etal-2019-opendialkg,tuan-etal-2019-dykgchat}, to ensure the coherence of the dialogue, the candidate topic set contains only the topics within one hop centered on the latest mentioned topic $e_c$, i.e., ${C}_t=\left \{ e|(e,r,e_c)\in \mathcal{G}  \right \}$.
The user utterance $u_{t}=\left \{(e_{t,1},p_{e_{t,1}}),...,(e_{t,\left | u_t \right | },p_{e_{t,\left | u_t \right | }})\right \}$ is the set of topic and preference pairs.


%% file: body/model.tex
\begin{figure}[t]
    \centering
    \includegraphics[height=0.15\textwidth]{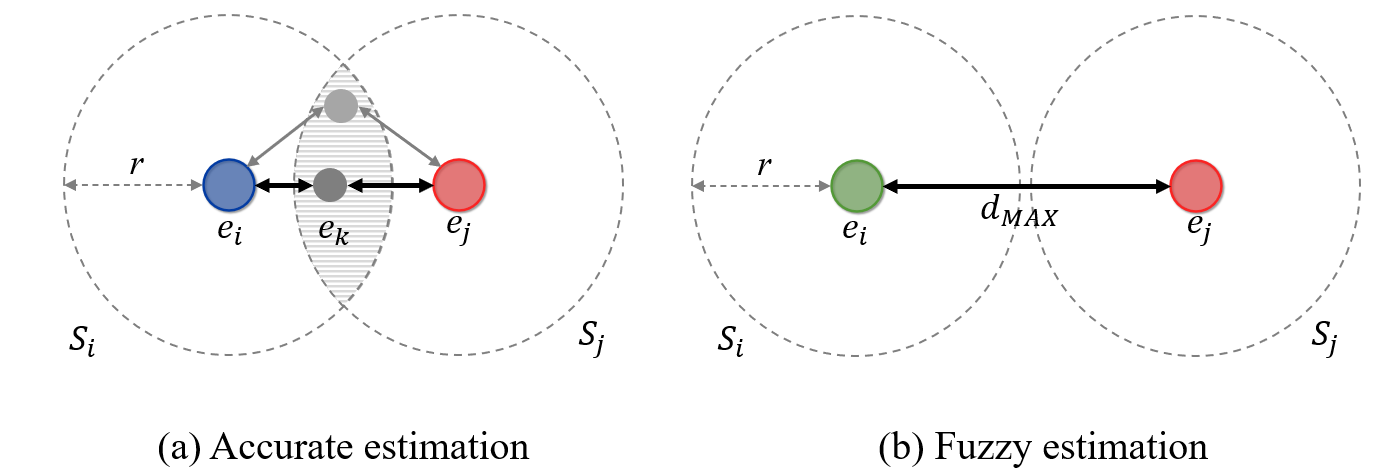}
    \caption{Soft distance estimation method. If the distance between two topics is less than the pre-set distance limit, we perform
    accurate distance estimation (a); otherwise, we perform fuzzy distance estimation (b).
    }
    \label{fig:edt} 
    \end{figure}

\section{Proactive Dialogue Policy}
\label{strategy}

With the proactive dialogue scenario setting, we propose a proactive dialogue policy \modelname{}, 
which employs a learned goal weight to achieve the desired trade-off between the targets of reaching the goal topic quickly, and maintaining a high user satisfaction.
At the $t$-th turn, with the goal topic $e_{g}$ and the dialogue history $H_{t-1}$ as input, \modelname{} chooses a topic $e_{t}^{a}$ from the candidate topic set~${C}_t$ as output.
During the topic choosing process, \modelname{} mainly tackles three issues: i) which topic is closest to the goal, ii) which topic the user prefers, and iii) how to make a trade-off between the two targets.
Accordingly, \modelname{} consists of a distance estimation module, a preference estimation module and a goal weight learning module.

\subsection{Distance Estimation}

In order to lead the conversation to the goal, it is inevitable to clarify which topic is closest to the goal.
The most straightforward method is to traverse the entire KG in advance to obtain the shortest distance between topics.
However, KGs are usually large-scale and continuously evolving, leading to high computational costs for global searches.
We propose a soft distance estimation method.
Specifically, if the distance between two topics is less than a pre-set distance limit $D$, we perform accurate distance estimation; otherwise, we perform fuzzy distance estimation.

We take the distance estimation between topics~$e_i$ and $e_j$ as an example to detail the soft distance estimation method, as shown in Figure~\ref{fig:edt}.
First, for the topic $e_i$ ($e_j$), we search with it as the center within radius $r$, where $r=\tfrac{1}{2} D$, and collect the topics covered during the search as the set $S_i$ ($S_j$).
Then, if the two sets, $S_i$ and $S_j$, intersect, we perform accurate distance estimation: using the minimum value of the sum of the distances from the overlapping topics to the two centers.
If the two sets are disjoint, we use a maximal value~$d_{MAX}$ as the fuzzy distance estimate.
The process can be formulated as:
\begin{equation}
    ed_{i,j}=\left\{\begin{matrix}
        \min_{e_k\in S_i\cap S_j} (d_{i,k}+d_{j,k}),& S_i\cap S_j \neq \varnothing \\
        d_{MAX}, &S_i\cap S_j = \varnothing 
      \end{matrix}\right.
\end{equation}
where  $ed_{i,j}$ is the estimated distance between $e_i$ and $e_j$,
$e_k$ belongs to the intersection of two sets $S_i$ and $S_j$, $d_{i,k}$ and $d_{j,k}$ are accuracy distances from the center topics, $e_i$ and $e_j$, to topic $e_k$.

\subsection{Preference Estimation}

Which topic the user prefers is an inevitable consideration for agents working to improve user satisfaction.
However, the agent can only obtain partially accurate preference information from the user through interaction.
Therefore, we propose a preference estimation method to estimate unknown preferences based on the known ones.

According to the core assumptions in recommender systems~\cite{su2009survey,Yue2014Collaborative,Dawen2016Factorization}, user preferences $P\in \mathbb{R} ^{\left | \mathcal{E}  \right |}$ can be viewed as the product of topic embeddings $E\in \mathbb{R} ^{\left | \mathcal{E}  \right | \times d}$ and the $d$-dimensional user vector $u\in \mathbb{R} ^d$, i.e., $P=Eu$.
We can obtain the user vector by solving this equation using the known preference information and its corresponding topics.
At the $t$-th turn, we estimate the user vector by using the normal equation:


\begin{figure}[t]
    \centering
    \includegraphics[height=0.15\textwidth]{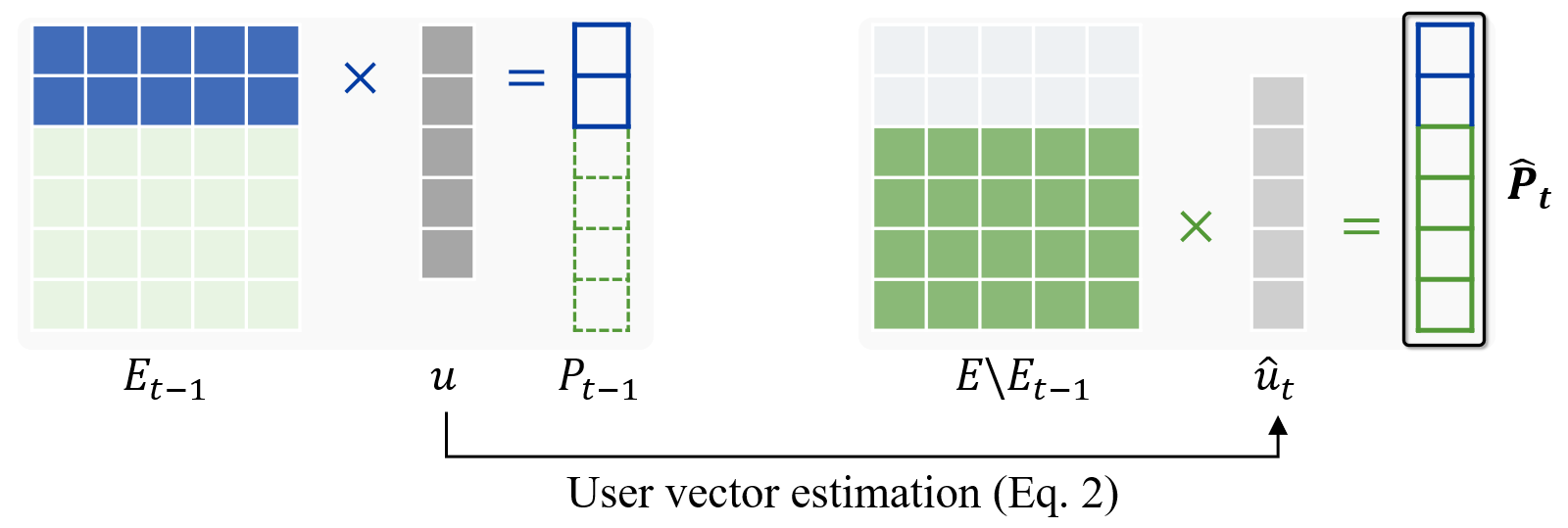}
    \caption{Preference estimation method. We first obtain the estimated user vector~$\hat{u}_t$ based on the known preference information and its corresponding topics. Then we use $\hat{u}_t$ to estimate the unknown user preferences and obtain the user preferences $\hat{P}_t$.
    }
    \label{fig:ept} 
    \end{figure}

\begin{equation}
    \hat{u}_t = ({E}_{t-1}^T {E}_{t-1}+\beta I)^{-1} {E}_{t-1}^T {P}_{t-1},
\end{equation}
where $\hat{u}_t$ is the estimated user vector~\footnote{Note that the estimated user vector calculated here will be used at the $t$-th turn, so we use $\hat{u}_t$ instead of $\hat{u}_{t-1}$.}, $E_{t-1}$ and $P_{t-1}$ are the topic embedding matrix and their corresponding user preference matrix for all topics that appear in the dialogue history $H_{t-1}$. The topic embeddings are learned by the node embedding learning model~\footnote{\url{https://github.com/thunlp/OpenNE}}. 
Considering the computational complexity for real-time interaction, we aware that Eq.~(2) is suitable for the case when the number of topics is small. When the number of topics is too large 
, we can use gradient descent to solve $\hat{u}_t$.

Finally, we use the estimated user vector to estimate the unknown user preferences and obtain the user preferences: 

\begin{equation}
    \hat{P}_t = P_{t-1}\oplus(E\setminus E_{t-1})~\hat{u}_t ,
\end{equation}
where the whole user preferences set $\hat{P}_t$ contains the accurate user preferences and estimated user preferences.

\subsection{Goal Weight Learning}

This module is proposed to achieve the trade-off between the target \uppercase\expandafter{\romannumeral1}) \textit{reaching the goal topic quickly} and target \uppercase\expandafter{\romannumeral2}) \textit{maintaining a high user satisfaction} to keep the user engaged during the conversation.
At the $t$-th turn, given the dialogue history $H_{t-1}$ and the goal topic $e_{g}$, the agent learns a goal weight~$gw_t$ to denote the importance of the target \uppercase\expandafter{\romannumeral1} and $1-{gw}_t$ to denote the importance of the target \uppercase\expandafter{\romannumeral2}.
Then, the weighted score of each topic in the candidate topic set~${C}_t$ is calculated:
\begin{equation}
    Score(e_{t,i})=gw_t\times Rank_d(ed_{i,g})+(1-gw_t) \times Rank(ep_{t,i}),
    \label{eq:score}
\end{equation}
where $e_{t,i}$ is the $i$-th topic in ${C}_t$, $ed_{i,g}$ is the estimated distance between $e_{t,i}$ and the goal topic $e_{g}$, $ep_{t,i} \in \hat{P}_t$ is the estimated user preference of $e_{t,i}$.
To eliminate the difference of magnitude between $ed_{i,g}$ and $ep_{t,i}$, we transform the raw $ed_{i,g}$ value into its descending ranking and the raw $ep_{t,i}$ value into its ascending ranking by the rank functions $Rank_d(\cdot )$ and $Rank(\cdot )$, respectively.
The topic with the highest score in ${C}_t$ will be chosen as $e_t^a$.

We abstract four factors from the dialogue history to blend into the goal weight $gw_t$.
The specific factors are: i) dialogue turn and ii) goal completion difficulty, related to the target \uppercase\expandafter{\romannumeral1};
iii) user satisfaction estimation and iv) cooperative degree, related to the target \uppercase\expandafter{\romannumeral2}.
Following are the details:

\begin{itemize}[leftmargin=*]
    \item[i)]\textbf{Dialogue turn $\bm t$} We expect the agent to lead the conversation to the goal as soon as possible. In that case, as the number of dialogue turns increases, so does the importance of achieving the goal. Thus, we introduce the dialogue turn $t$ vector in the goal weight.

    \item[ii)]\textbf{Goal completion difficulty $\bm {gcd_t}$}
    Another factor related to the target \uppercase\expandafter{\romannumeral1} is the difficulty of completing it. 
    We approximate this difficulty by calculating the distance between the current talked topic $e_c$ and the goal topic $e_g$, i.e., $gcd_t=ed_{c,g}$.
    A high~$gcd_t$ indicates that the current conversation is still far from the goal and that it will take a lot of effort to lead the conversation to the target.
  Conversely, the agent can easily complete the target \uppercase\expandafter{\romannumeral1}.

    \item[iii)]\textbf{User satisfaction estimation $\bm {eus_t}$}
    This factor directly portrays the time-realistic state of the target \uppercase\expandafter{\romannumeral2}. 
    We estimate it by using information about user preferences that appears in user utterances.
    Specifically, we calculate the average preference value of all user utterances in the previous $t-1$ turns to yield the estimated user satisfaction, i.e., $eus_t=\textstyle \frac{1}{t-1}{\textstyle \sum_{i=1}^{t-1}\frac{1}{\left | u_i \right | }  {\textstyle \sum_{j=1}^{\left | u_i \right | }ep_{e_{i,j}}} }$, where $e_{i,j}\in u_{i}$ denotes the $j$-th topic mentioned by user at the $i$-th turn, and $ep_{e_{i,j}}\in\hat{P}_t$ denotes the preference of $e_{i,j}$.

    \item[iv)]\textbf{Cooperative degree $\bm {cd_t}$}
This factor is learned from the historical user behaviors.
At the $t$-th turn, we generate a one-hot sequence based on the user behavior in the previous $t-1$ turns, where 0 indicates that the user cooperate in the current turn, 1 indicates the user did not cooperate.
Then we use GRU\cite{cho2014properties} to encode this one-hot sequence to yield the cooperative degree~$cd_t$.
When $cd_t$ is low, the user is easy to behave uncooperatively, so the agent should be careful to avoid triggering uncooperative user behavior.


    
\end{itemize}

Finally, we concatenate four factors and adopt a 2-layer policy MLP to learn the goal weight:
\begin{equation}
    gw_t  =MLP( t\oplus gcd_t \oplus eus_t \oplus cd_t ).
    \end{equation}

\subsection{Deep Q-Network Learning}

We model the process by which an agent chooses a new topic based on a dialogue policy as a Markov Decision Process (MDP).
In details, as a RL problem, at the $t$-th turn, $\left \langle s_{t}, a_{t}, p_{t}, r_{t}\right \rangle$ in MDP are defined as: 
1) State $s_{t}$ is equal to the dialogue history~$H_{t-1}$ with the goal topic $e_{g}$;
2) Action $a_{t}$ is equal to the candidate topic set~${C}_t$;
3) Transition $p_{t}$ is the transition function with $p(s_{t+1}|s_{t},e_{t}^{a})$ being the probability of seeing state~$s_{t+1}$ after taking action $e_{t}^{a}$ based on~$s_{t}$;
and 4) Reward $r_{t}$ is the reward that the agent obtains from the environment, including rewards reflecting changes in goal completion and user satisfaction.
Our goal is to learn a dialogue policy which can maximize the cumulative reward over the whole conversation process as:
\begin{equation}
    \theta ^{*}=\arg \max_{\pi \in \prod }\mathbb{E}\left [ \sum_{t=1}^{T}r(s_{t},e_{t}^{a})\right ],
\end{equation}
where $\theta ^{*}$ is the optimal parameter set of \modelname{} and $r_{t}=r_{us}+r_{quit}+r_{goal}+r_{suc}+r_{fail}$.
Specifically, 
$r_{us}$ is the change of the user satisfaction, i.e., $\alpha(US_{t}-US_{t-1})$, where $\alpha$ is the magnification factor;
$r_{quit}$ is a strongly negative reward if the user quits the conversation;
$r_{goal}$ is the change of the goal completion degree, i.e., $e^{-\lambda t}(d_{t}-d_{t-1})$, where $\lambda$ is the time decay coefficient, $d_{t}$ denotes the distance between the current topic at the $t$-th turn and the goal topic;
$r_{suc}$ is a strongly positive reward if the conversation reach the goal topic;
and $r_{fail}$ is a strongly negative reward if the conversation do not reach the goal topic before the maximum turn limit $T$.

We adopt Q-learning algorithm to train \modelname{}.
We use the score calculated by Eq.~\ref{eq:score} as the Q-value~$Q_{\theta} (s_{t},e_{t}^{a})$.
Based on the optimal Bellman Equation~\cite{1998Reinforcement}, we can represent the optimal Q-value with the the maximum expected reward achievable is:

\begin{equation}
    Q_{\theta}^{\ast} (s_{t},e_{t}^{a})=\mathbb{E}\left [ r_t + \gamma \operatornamewithlimits{max}_{e_{t+1}^{a}\in a_{t+1}}Q_{\theta}(s_{t+1},e_{t+1}^{a})\right ].
    \label{eq:yt}
\end{equation}

We improve the value function $Q_{\theta}(s_{t},e_{t}^{a})$ by adjusting $\theta$ to minimize the mean-square loss function, defined as follows:
\begin{equation}
    l(\theta) = \mathbb{E}_{(s_t,a_t,s_{t+1},r_t)\sim {M}}[(Q_{\theta}^{\ast} (s_{t},e_{t}^{a})-Q_{\theta}(s_t,a_t))^2].
    \label{eq:loss}
\end{equation}

%% file: body/experiment.tex
\section{Experiments}
\label{experiments}

The key contributions of this work are on studying the proactive dialogue policy interactively and the design of a proactive dialogue policy \modelname{}.
We first conduct a comparison experiment to evaluate the effectiveness of \modelname{} interactively.
Then we explore the impact of different types of users on the proactive dialogue policy.
In addition, we perform ablation study to investigate the effect of each factor in the goal weight learning.
Finally, we conduct a case study to demonstrate the superiority of \modelname{}.
Specifically, we have the following research questions (RQs) to guide experiments:
\begin{itemize}[leftmargin=*]
    \item[\small] \textbf{RQ1:} How is the overall performance of \modelname{} comparing with existing dialogue policies?
    
    \item[\small] \textbf{RQ2:} How do different types of users affect the proactive dialogue policy?
    
    \item[\small] \textbf{RQ3:} How do different factors affect the goal weight learning?
    \end{itemize}


\begin{table*}[t]
    \renewcommand\arraystretch{1.3}
    \footnotesize      
        \begin{center}
        \caption{
            GCR(\%) and US(\%) performance~($\pm$ standard deviation) of compared dialogue policies~(pairwise t-test at 5\% significance level). 
            All dialogue policies are evaluated on three types of user simulators with different tolerance values~($k=0.8$, $1.0$ and $1.2$) as well as on a mixed-user simulator. 
        }
        \label{table:mainresult} 
        \resizebox{0.98\textwidth}{!}{
        \begin{tabular}{lrr|rr|rr|rr}
        \toprule[1.2pt]
        \multirow{2}{*}{}  & \multicolumn{2}{c|}{Mixed User Simulator} & \multicolumn{2}{c|}{User Simulator ($k=0.8$)}  & 
        \multicolumn{2}{c|}{User Simulator ($k=1.0$) }    & 
        \multicolumn{2}{c}{User Simulator ($k=1.2$) }\\ \cline{2-9}\rule{0pt}{8pt}
    & \multicolumn{1}{c}{GCR} &   \multicolumn{1}{c|}{US}  & \multicolumn{1}{c}{GCR} &   \multicolumn{1}{c|}{US}  & \multicolumn{1}{c}{GCR} &   \multicolumn{1}{c|}{US}  & \multicolumn{1}{c}{GCR} &   \multicolumn{1}{c}{US}    \\\midrule[1.2pt]
    Random  &		1.29\scriptsize{~$\pm$0.01}&		52.86\scriptsize{~$\pm$0.01} & 1.67\scriptsize{~$\pm$0.01}&		55.29\scriptsize{~$\pm$0.01}&		1.18\scriptsize{~$\pm$0.01}&		52.65\scriptsize{~$\pm$0.01}&		1.02\scriptsize{~$\pm$0.01}&		50.65\scriptsize{~$\pm$0.01}	\\

    Pop~(GCR) &	57.00\scriptsize{~$\pm$0.03}&	53.52\scriptsize{~$\pm$0.00} & 48.18\scriptsize{~$\pm$0.03}&	56.05\scriptsize{~$\pm$0.00}&	59.18\scriptsize{~$\pm$0.03}&	53.24\scriptsize{~$\pm$0.00}&	63.63\scriptsize{~$\pm$0.02}&	51.27\scriptsize{~$\pm$0.00}  \\

    Pop~(US)  & 1.16\scriptsize{~$\pm$0.01}&	62.14\scriptsize{~$\pm$0.00} & 1.39\scriptsize{~$\pm$0.01}&	64.47\scriptsize{~$\pm$0.00}&	1.07\scriptsize{~$\pm$0.01}&	61.80\scriptsize{~$\pm$0.00}&	1.03\scriptsize{~$\pm$0.00}&	60.15\scriptsize{~$\pm$0.00}\\\midrule[0.9pt]
    
    NICF &	3.46\scriptsize{~$\pm$0.01}&	54.48\scriptsize{~$\pm$0.00} &	3.55\scriptsize{~$\pm$0.01}&	57.76\scriptsize{~$\pm$0.00}&	3.30\scriptsize{~$\pm$0.01}&	54.10\scriptsize{~$\pm$0.00}&	3.52\scriptsize{~$\pm$0.01}&	51.58\scriptsize{~$\pm$0.01}	\\

    DeepPath  &	3.84\scriptsize{~$\pm$0.01}&	54.41\scriptsize{~$\pm$0.00}& 4.32\scriptsize{~$\pm$0.01}&	57.94\scriptsize{~$\pm$0.00}&	3.81\scriptsize{~$\pm$0.01}&	54.06\scriptsize{~$\pm$0.00}&	3.38\scriptsize{~$\pm$0.01}&	51.24\scriptsize{~$\pm$0.00}\\

    NKD &	3.15\scriptsize{~$\pm$0.01}	&	52.96\scriptsize{~$\pm$0.00}	&	4.19\scriptsize{~$\pm$0.02}	&	56.45\scriptsize{~$\pm$0.01}	&	2.97\scriptsize{~$\pm$0.01}	&	52.31\scriptsize{~$\pm$0.00}	&	2.28\scriptsize{~$\pm$0.01}	&	50.13\scriptsize{~$\pm$0.00}	\\\midrule[0.9pt]

    \modelname{}  &	\textbf{29.74}\scriptsize{~$\pm$0.07}&	\textbf{59.28}\scriptsize{~$\pm$0.01}	&	\textbf{22.47}\scriptsize{~$\pm$0.06}	&	\textbf{61.91}\scriptsize{~$\pm$0.01}	&	\textbf{32.20}\scriptsize{~$\pm$0.07}	&	\textbf{58.64}\scriptsize{~$\pm$0.01}	&	\textbf{34.54}\scriptsize{~$\pm$0.08}	&	\textbf{57.28}\scriptsize{~$\pm$0.01}	\\\bottomrule[1.2pt]

        \end{tabular}
        }
        \end{center}
        \end{table*}

\subsection{User Simulation}
\label{user}

Considering the huge costs associated with involving humans in the interaction, we design user simulators to complete the process of interacting with the agents when conducting the experiments.
The design of the simulator is also at the topic-level.
When interacting with the agent, at the $t$-th turn, the simulator accepts the dialogue history and the topic introduced by the agent at the current turn, i.e., $H_{t-1}\cup \left \{ e_{t}^{a} \right \}$. 
Then, it responds $u_{t}$ to the agent or quits the conversation.

Here, we describe in detail how to generate $u_{t}$.
The user utterance is based on satisfaction with the current conversation.
We argue that satisfaction can be formalized as the cumulative average of users' preferences for the topics covered by the conversation:
\begin{equation}
    US_{t}\triangleq {\textstyle \frac{1}{t}\sum_{i=1}^{t} {\textstyle \frac{1}{\left | u_i+1 \right | } (\sum_{j=1}^{\left | u_{i} \right |}p_{e_{i,j}}+p_{e_i^a})} },
\end{equation}
where $US_{t}$ denotes the user satisfaction at the $t$-th turn, $e_{i,j}\in u_{i}$ denotes the $j$-th topic mentioned by user at the $i$-th turn and $p_{e_{i,j}}$ denotes the user preference to the topic $e_{i,j}$.
${\left | u_i \right | }$ is the total number of topic mentioned in the $t$ turn utterance (without removing duplicates).
Note that when $e_{i,j} =  e_{i}^{a}$, there is no need to double count the latter $e_{i}^{a}$.
User preferences are personalized. 
For one user simulator, we first sample a user vector~$u\in \mathbb{R} ^d$.
Each element in the user vector is sampled from the Gaussian distribution $N(0,2)$.
Next, following the core assumptions in recommendation techniques~\cite{su2009survey,Yue2014Collaborative,Dawen2016Factorization}, we map the user vector to the topic embeddings $E\in \mathbb{R} ^{\left | \mathcal{E}  \right | \times d}$ to calculate user preferences, i.e., $P=Eu$, where $P=\left \{ p_e|e\in\mathcal{E} \right \} $.

At the $t$-th turn, based on the calculated user satisfaction, the user utterance can be generated in cases:
\begin{equation}
        u_{t}=\begin{cases}
            \left \{(e_{t}^{a},p_{e_{t}^{a}})\right \} & \text{ if }  US_{t}> Q_c, \\
               \left \{(e,p_{e})\mid e\in \mathcal{E} \wedge  Sel(e)=1\right \}& \text{ if }Q_q < US_{t}\le Q_c, \\
               \varnothing  & \text{ if } US_{t}\le Q_q,
              \end{cases}
\end{equation}
specifically, the user behavior can be deconstructed into three types: cooperative, non-cooperative and quit.
If $US_{t}$ is above the cooperative threshold $Q_c$, the simulator will only respond the topic introduced by agent at this turn and the corresponding preference.
Whereas if $US_{t}$ is below the cooperative threshold $Q_c$, the simulator will respond new topics with their preferences.
The new topic in $u_{t}$ is chosen with its user preference as probability.
We design the function $Sel(e)$ to calculate whether the topic $e$ is chosen or not:
\begin{equation}
    Sel(e) \sim  Bernoulli(p_e),
\label{eq:selection}
\end{equation}
where the $Sel(e)$ is sampled from the Bernoulli distribution: 1 indicates the topic is chosen, while 0 indicates the opposite.
If $US_{t}$ fall below an even lower quit threshold $Q_q$, the simulator will quit the conversation in advance.

Finally, to further imitate humans, we personalize the thresholds $Q_c$ and $Q_q$ as well.
We design a new characteristic ``tolerance''.
A high tolerance simulator is less likely to behave non-cooperatively or to quit a conversation outright, while a low tolerance simulator does the opposite.
Concretely, the tolerance character $k$ is used as a shrinker on the thresholds.
That is, we use $\tfrac{1}{k} Q_c^*$ and $\tfrac{1}{k} Q_q^*$ as the personalized cooperative and quit thresholds, where $Q_c^*$ and $Q_q^*$ are pre-set constants.


In the evaluation process, for one start-goal topic pair, we design three user simulators with the same user vector and different tolerance characteristics, low~($k=0.8$), medium~($k=1.0$) and high~($k=1.2$), respectively.
We choose 0.5 as~$Q_c^*$ and 0.4 as~$Q_q^*$, experimentally.
Considering the computational consumption caused by large-scale KG, we set the topic selection scope of the simulator to all neighbors within 3 hops of the current topic, while $\left | u_t \right | $ is limited to maximum 3.

\subsection{Settings}
\subsubsection{Dataset}
\label{dataset}
The dataset we used is collected from KdConv~\cite{zhou2020kdconv}, which contains a film-related KG and a large amount of multi-turn conversation data.
First, we continue to use the KG of KdConv, which has 7477 entities, 4939 relations and 89618 triples.
Then, because this work focuses on studying the dialogue policy in an interactive manner, we abandon the static conversation data, and extract only the start-goal topic pairs from the multi-turn conversation data.
Specifically, the start topic is extracted from the $1^{st}$ turn and the goal topic is extracted from the last turn of the conversation.
Statistically, the KG has 7477 film-related topics.
The training and testing sets contain 353 and 89 start-goal topic pairs, respectively.

\subsubsection{Evaluation Metrics}
\label{metrics}

With the targets \uppercase\expandafter{\romannumeral1} and \uppercase\expandafter{\romannumeral2}, we use two metrics to automatically evaluate the all-turn performance of the dialogue policy:

\begin{itemize}[leftmargin=*]

\item[\small] \textbf{Goal Completion Rate~(GCR)} We use GCR to evaluate the ability of the policy to lead the conversation successfully. GCR is defined as: 
\begin{equation}
\mathrm{GCR} =\textstyle\frac{1}{N}\textstyle \sum_{k=1}^{N}  e^{-\lambda T_k} D_k
\end{equation}
where $N$ is the total number of dialogues, $T_k$ is the total number of turns of the $k$-th dialogue, $\lambda$ is the time decay coefficient. $D_k$ is set to 1 if the $k$-th dialogue succeeded in achieving the agent's goal, otherwise it is 0.

\item[\small] \textbf{User Satisfaction~(US)} We use US to evaluate the ability of the policy to maintain the high user satisfaction. 
We use the average preference of all the topics involved in the whole conversation to represent US:
\begin{equation}
    \mathrm{US}=\textstyle\frac{1}{N}\sum_{k=1}^{N}{\textstyle \frac{1}{T_k}\sum_{i=1}^{T_k}\frac{1}{\left | u_i \right | +1}( {\textstyle \sum_{j=1}^{\left | u_{k,i} \right |}p_{e_{k,i,j}}} +p_{e_{k,i}^{a}}}),
\end{equation}
where $p_{e_{k,i,j}}\in u_{k,i}$ is the preference of the $j$-th topic mentioned by user at the $i$-th turn in the $k$-th dialogue, and $p_{e_{k,i}^{a}}$ is the preference of the topic introduced by the agent at the $i$-th turn in the $k$-th dialogue. ${T_k}$ is the total number of dialogue turns for the $k$-th conversation. Note that when $e_{k,i,j} =  e_{k,i}^{a}$, there is no need to double count the latter $e_{k,i}^{a}$.
\end{itemize}

\subsubsection{Training Details}
\label{training}
The dimensions of the user vector and topic embedding are all set to $50$.
The node embedding learning model is DeepWalk~\cite{perozzi2014deepwalk}.
The maximum dialogue turn $T$ is set to $20$. 
The distance limit $D$ used in the distance estimation is set to 6.
The optimizer chosen is the Adam and we empirically set the learning rate as $1e{-7}$.
The maximum training epoch is set to 100.
The RL algorithm used here is DQN\cite{mnih2015human}, in which the memory size, batch size, decay factor $\gamma$, exploration factor $\varepsilon$ for the $\varepsilon$-greedy policy, time decay coefficient~$\lambda$, and the magnification factor $\alpha$ are $2000$, $100$, $0.9$, $0.2$, $0.02$ and $100$, respectively. 
Rewards $r_{quit}, r_{suc}, r_{fail}$ are $-10$, $20$ and $-10$ if triggered, respectively; otherwise they are all $0$.


For an in-depth and extensive evaluation, we test all comparison dialogue policies for 100 rounds.
Due to a certain randomness of user behavior, there is a slight variation in the results of 100 rounds.
We expect to select the best result for reporting.
However, under the multi-metric evaluation setting, the two metrics we use (GCR and US) are equally important, so the best results are often a set.
In this set, no one result can outperform the other results on both metrics at the same time.
Therefore, we report the average of all results in the optimal results set as the final performance for each policy.


\subsubsection{Baselines}
\label{baseline}

This work is the first attempt to study a proactive dialogue policy which has the ability to quickly achieve the goal topic as well as maintain high user satisfaction.
We first present three baselines to perceive the difficulty of this task: 
\begin{itemize}[leftmargin=*]
\item[\small] \textbf{Random} At each turn, the agent randomly selects the topic.
\item[\small] \textbf{Pop(GCR)} At each turn, the agent chooses the topic closest to the goal, i.e., with the smallest estimated distance, in order to get the highest GCR.
\item[\small] \textbf{Pop(US)} At each turn, the agent chooses the user's favorite topic, i.e., with the highest estimated preference, in order to get the highest US.
\end{itemize}

\begin{figure}[t]
    \centering
    \includegraphics[height=0.23\textwidth]{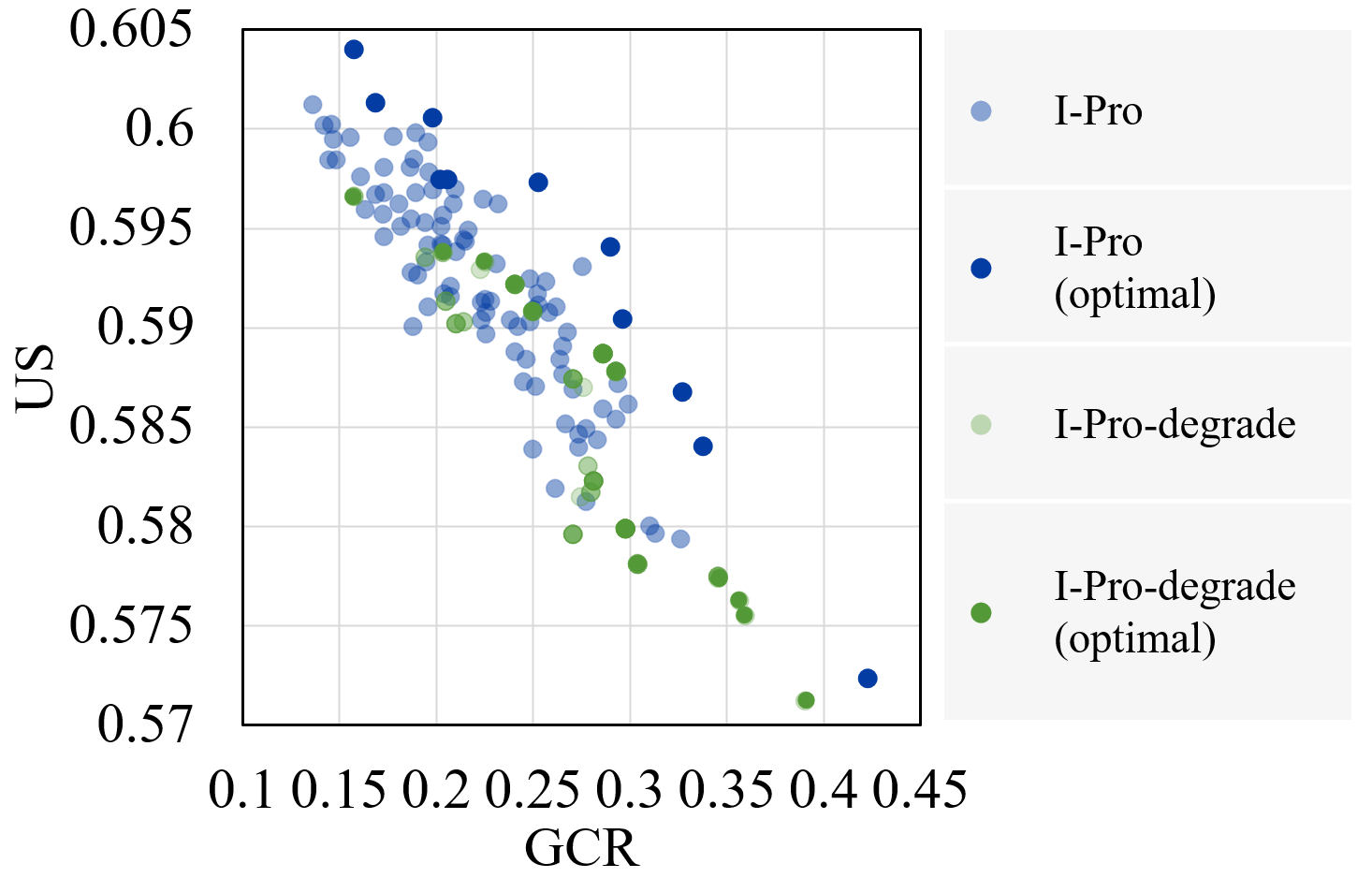}
    \caption{Comparison performance of \modelname{} and the Variant \modelname{}-degrade. 
    The optimal results for \modelname{} and \modelname{}-degrade are depicted.
    Apparently, the optimal results of \modelname{} completely surround the optimal results of \modelname{}-degrade.
    }
    \label{fig:degrade} 
    \end{figure}

\begin{figure}[t]
    \centering
    \includegraphics[height=0.23\textwidth]{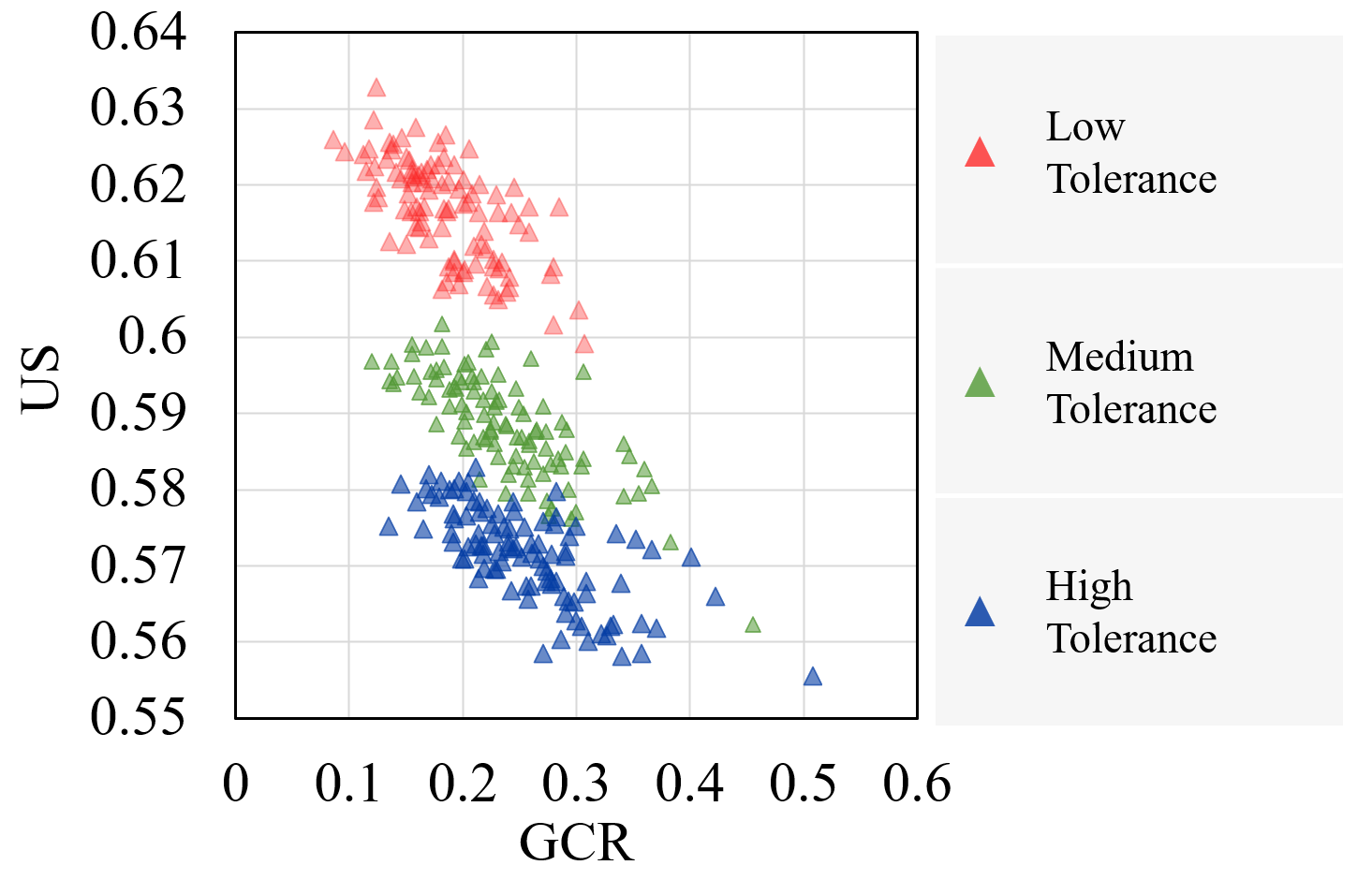}
    \caption{Performance w.r.t. different user simulator types.
    Performance on three user simulator types with different tolerance: low~($k=0.8$), medium~($k=1.0$) and high~($k=1.2$) are visualized.
    GCR increases and US decreases as the tolerance level increases.
    }
    \label{fig:user} 
    \end{figure}

Considering that there are currently no dialogue policies that address both targets, we chose several policies that are committed to one target alone as the baselines.
\begin{itemize}[leftmargin=*]
    \item[\small] \textbf{NICF}~\cite{NICF} This model is originally designed to recommend items for user in an interactive setting. In our proactive dialogue scenario, it can be used to enhance US during multi-turn interactions.

    \item[\small] \textbf{DeepPath}~\cite{wenhan_emnlp2017} This model is focus on searching a path in KG through multi-hop reasoning from the start entity to the goal entity. Here, it can work on leading the conversation to the goal topic.
    
    \item[\small] \textbf{NKD}~\cite{liu-etal-2018-knowledge} This model works on multi-turn dialogue generation based on the dialogue history and KG.
    We degrade it to the topic level dialogue generation.
\end{itemize}

All the above baselines and our \modelname{} are trained and tested in an interactive manner with our designed user simulators.

\subsection{Overall Performance (RQ1)}
\label{exp1}


\noindent
\textbf{\modelname{} vs. Baselines}
Table~\ref{table:mainresult} demonstrates the performance of \modelname{} and baselines. 
As can be clearly seen, \modelname{} significantly outperforms existing dialogue policies both on GCR and US.
We also make the following observations:
1) All policies show a decrease in the US metric as tolerance increases, because users with low tolerance are more likely to be triggered to quit the conversation at a higher satisfaction level.
2) The Random baseline shows a decreasing trend under the GCR metric as user tolerance increases. When the user tolerance is low, there is more non-cooperative behavior from the user.
And the perturbation caused by the user's non-cooperative behavior on the conversation forward direction makes the conversation have a chance to reach the goal topic.
3) Two ``Pop'' baselines of our own design show the upper bounds of the effects currently achievable on both GCR and US metrics.
These two upper bounds can be further improved by improving the estimation accuracy of user preferences and the distance.
Interestingly, under the GCR metric, the Pop~(GCR) tends to increase as tolerance increases, while the Pop~(US) does the opposite.
We speculate that this is because the Pop~(GCR) aims to improve GCR, so the lower the user tolerance, the more likely it is to trigger non-cooperative behavior.
While, Pop~(US), same as Random, has a higher probability of the agent walking to the goal by chance at low user tolerance.
4) Three existing works, NICF~\cite{NICF}, DeepPath~\cite{wenhan_emnlp2017} and NKD~\cite{liu-etal-2018-knowledge}, all underperform, exposing the challenges in proactive dialogue policy study.
We suspect that their drawbacks are the inaccurate estimation of user preferences and distance to the goal, as well as a lack of ability to make the trade-off between target \uppercase\expandafter{\romannumeral1} and \uppercase\expandafter{\romannumeral2}.

\begin{table}[t]
    \centering
    \normalsize
    \caption{Performance (GCR and US) and the learned goal weight value~($\pm$ standard deviation) at different tolerance values.}
    \label{tab:tolerance}
\resizebox{0.46\textwidth}{!}{
    \begin{tabular}{lccccccc}
        \toprule[1.2pt]
     k&0.4  &0.6  &0.8  &1.0  &1.2  &1.4  &1.6   \\\midrule[1.2pt]
     GCR&15.84  & 17.04 & 22.47 & 32.20 &34.54  & 34.1 & 34.75  \\\midrule[0.3pt]
     US& 62.37 & 62.71 & 61.91 & 58.64 & 57.28 & 56.24 & 55.99 \\\midrule[0.8pt]
     $\overline{gw}$ & 0.37 & 0.41  & 0.45 &	0.49 &	0.50 
       & 0.51 & 0.51  \\\midrule[0.3pt]
       $\pm$ sd& 0.17 & 0.17 & 0.17& 	0.13 &	0.11 
     & 0.10 & 0.09  \\\midrule[1.2pt]
    \end{tabular}}
    \end{table}

\noindent
\textbf{\modelname{} vs. \modelname{}-degrade} 
We propose one variant of \modelname{}, namely {\modelname{}-degrade}, to further explore the utility of our proposed goal weight.
In this variant, the goal weight is degraded to one trainable parameter, i.e., $Score(e_{t,i})=\beta\times Rank_d(ed_{i,g})$ $+(1-\beta) \times Rank(ep_{t,i})$. 
We intuitively present the performance comparison in Figure~\ref{fig:degrade}.
Clearly, the optimal results of \modelname{} completely surround the optimal results of {\modelname{}-degrade}.
This exhibits the shallow capability of one parameter to make the trade-off between two targets. 
We also observe that the effect of {\modelname{}-degrade} has many repeated values, which shows the inflexible ability of only using one parameter.
Apparently, it is intractable for one parameter to perceive the progress of the conversation and the personality of the user.
This suggests that our integration of multiple factors on the goal weight learning is necessary.

\subsection{Performance w.r.t. Different User Types (RQ2)}
\label{exp2}

To investigate the differences in the effects of \modelname{} on different types of users, we evaluate \modelname{} on three user simulator types with different tolerance: low~($k=0.8$), medium~($k=1.0$) and high~($k=1.2$), and visualize the results in Figure~\ref{fig:user}.
Obviously, the difference in tolerance leads to a significant difference in performance: GCR increases and US decreases as the tolerance level increases.
Users with low tolerance are more likely to behave non-cooperatively when their satisfaction is low, making it difficult for the agent to achieve the goal topic. In addition, agents chooses users' preferred topics in order to try to avoid non-cooperative user behaviors. Therefore, low tolerance users will lead to a low GCR and a high US.
High tolerance users are just the opposite.

We further explore more diverse tolerance values and the correlation between the tolerance values and the learned goal weight.
Here, for the goal weights, we use their averages over all dialogue turns.
Results are reported in Table~\ref{tab:tolerance}.
We observe that the goal weight is perceptible to changes in tolerance and there is a positive correlation between the two terms.
When the goal weight perceives that the user has a high tolerance, it will spontaneously be more biased towards improving GCR.
We also observe that the standard deviation of the goal weight is negatively correlated with the tolerance.
This is because the less tolerance the user has, the more drastic the change in user behavior (cooperative or non-cooperative) will be.
The goal weight reacts to the perceived change as a larger standard deviation.

\begin{figure}[t]
    \centering
    \includegraphics[height=0.23\textwidth]{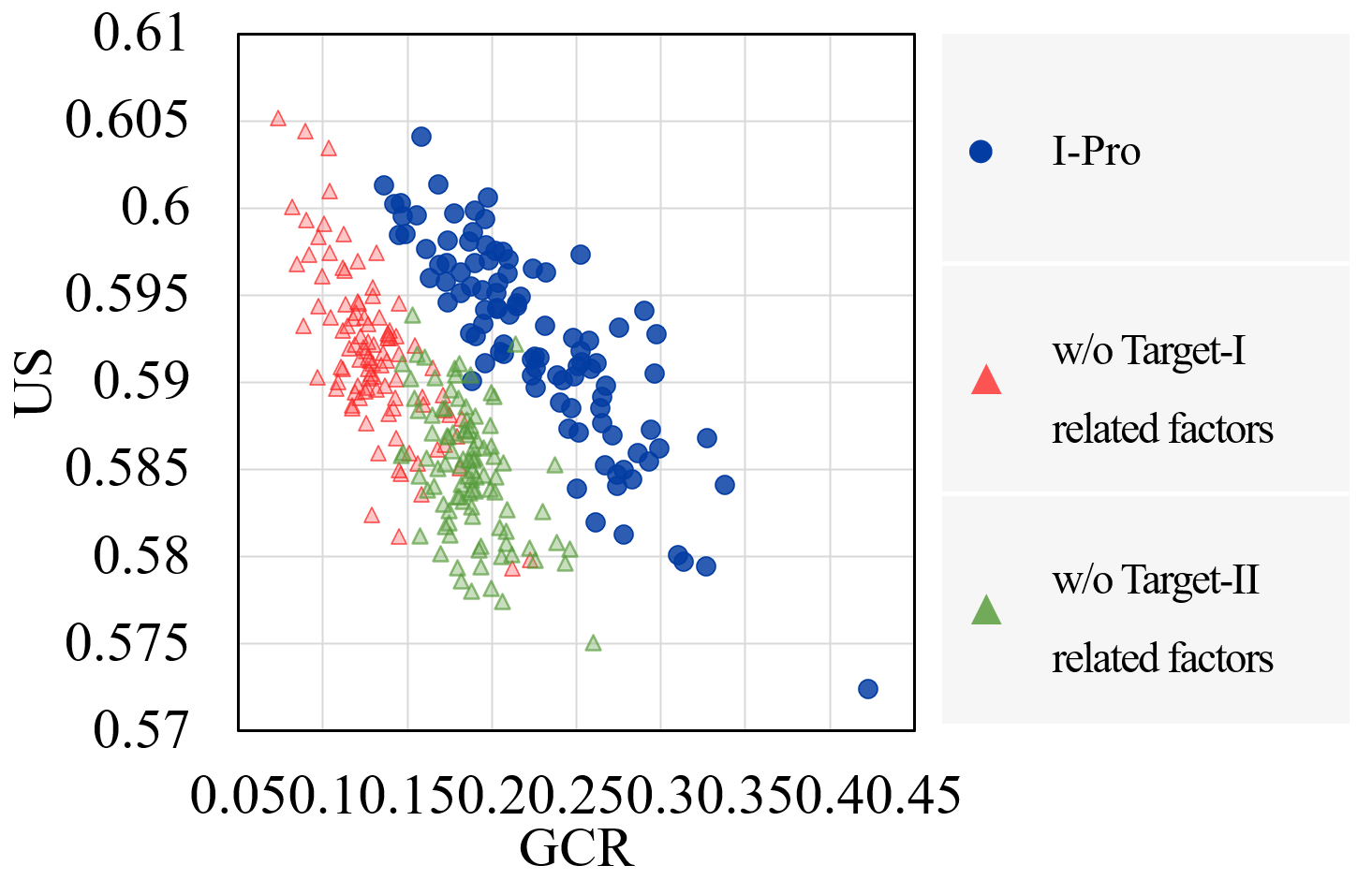}
    \caption{Ablation study I: performance on deactivating “approach the goal” related factors and “satisfy the user” related factors, respectively.}
    \label{fig:ablation1} 
    \end{figure}

\subsection{Ablation Study on Goal Weight (RQ3)}
\label{exp3}
We abstract four factors from the dialogue history to learn the goal weight.
Here, we first analyze the effectiveness of different factors in goal weight learning.
As shown in Figure~\ref{fig:ablation1}, we evaluate the performance on our two variants: 1) the goal weight learning without two target \uppercase\expandafter{\romannumeral1}-related factors: dialogue turn factor and goal completion degree factor, and 2) the goal weight learning without two target \uppercase\expandafter{\romannumeral2}-related factors: user satisfaction estimation factor and cooperative degree factor.
We observe that, when deactivating the target \uppercase\expandafter{\romannumeral1}-related factors, the GCR score decreases significantly and the US score increases slightly.
This confirms that when the agent is unable to perceive the progress of the conversation, its control over the conversation decreases, and the conversation is easily led to the user's preferred topic by the user's non-cooperative behaviors. 
This eventually manifests as an increase in user satisfaction, but the conversation fails to achieve the goal topic.
When deactivating the target \uppercase\expandafter{\romannumeral2}-related factors, both the US and GCR scores decrease.
This suggests that the conduct of proactive conversations is sensitive to user satisfaction and behavior.
So, it is critical to estimate the satisfaction of current users as well as to predict the cooperative degree of users. 
When deactivating the user satisfaction estimation factor, the agent cannot perceive user dissatisfaction in time, leading to a decrease in US. 
In addition, when the agent cannot perceive the cooperative degree of users, more non-cooperative user behaviors are triggered, leading to a decrease in GCR.

  \begin{figure}[t]
    \centering
    \includegraphics[width=0.43\textwidth]{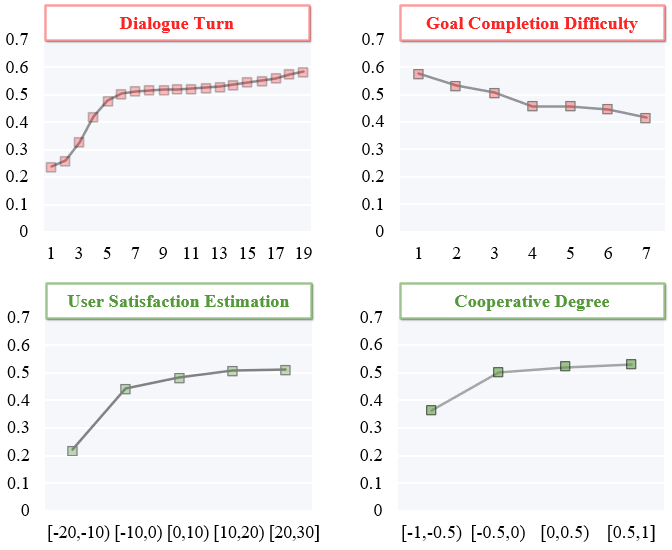}
    \caption{Ablation study II: correlations between different factors and the goal weight. The horizontal axis indicates the individual factors. The vertical axis indicates the corresponding goal weight values under the particular factor.}
    \label{fig:ablation2} 
    \end{figure}

We secondly analyze the specific correlation between each factor and the goal weight.
The correlations are pulled from the dialogue data collected during the evaluation, which are visualized in Figure~\ref{fig:ablation2}.
We make the following observations:

  \begin{figure*}[t]
    \centering
    \includegraphics[width=0.92\textwidth]{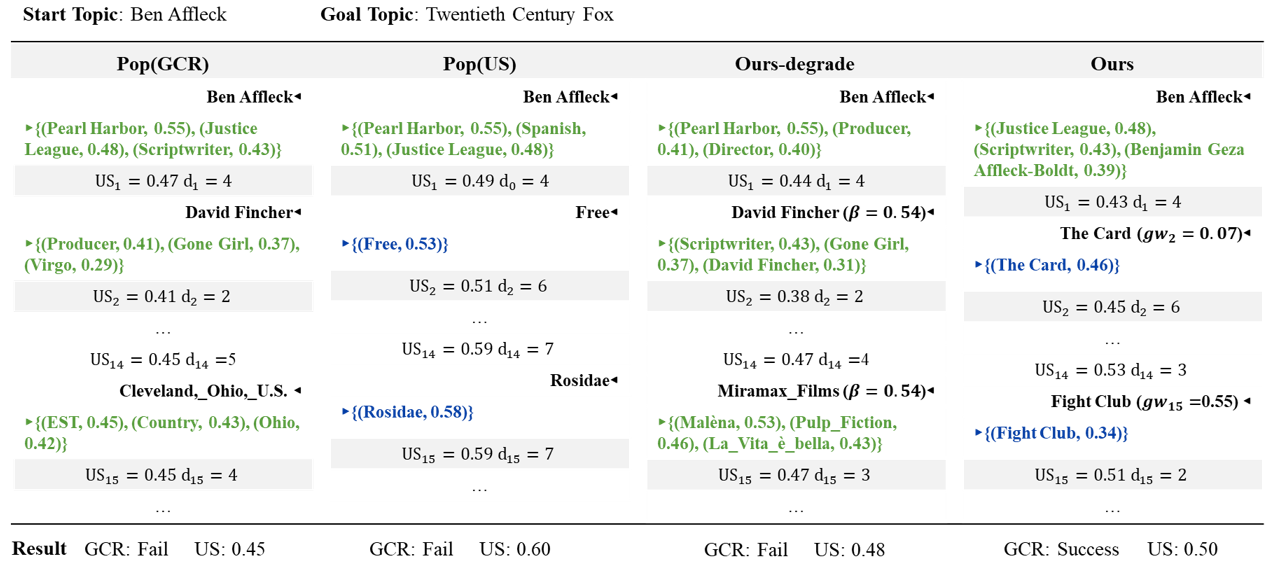}
    \caption{Case study of four dialogue policies: Pop (GCR), Pop (US), \modelname{}-degrade and \modelname{}. We choose the same start and goal topics, and record the interaction process between each dialogue policy and the user simulator whose tolerance is 1. We report partial transcripts of the conversations, including turns 1, 2, 14, and 15. Specifically, at each turn, we report the conversation detail, the current user satisfaction ($US_t$) and the distance to goal ($d_t$). The finally results also are reported. Specially, we also report the goal weight values of \modelname{}-degrade and \modelname{}, i.e., $\beta$ and $gw_t$.
    Due to the space limitation, we show here only the English translation of the cases.
    }
    \label{fig:case} 
    \end{figure*}

\begin{itemize}[leftmargin=5mm]
    \item[i)] The dialogue turn factor correlates positively with the goal weight.
    The agent aims to achieve the goal topic as soon as possible, thus, the goal weight increases with the number of dialogue turns. 
    Interestingly, we observe that the weight values are particularly low at the beginning of the conversation, which means that the agent is more willing to satisfy the user at the early stage.
    \item[ii)] The goal completion difficulty factor correlates negatively with the goal weight.
    Note that larger factor value indicates greater distance from the goal, i.e. higher goal completion difficulty.
    This is reasonable because when the agent perceives that the current conversation is close to the goal topic, the potential gain obtained by choosing a topic close to the goal is higher than that preferred by the user.
    \item[iii)] The user satisfaction estimation factor correlates positively with the goal weight.
    This factor directly reflects user satisfaction, so when the agent estimates that user satisfaction is low, it will be more inclined to choose topics that users prefer.
    \item[iv)] The cooperative degree factor correlates positively with the goal weight.
    This factor models the tolerance character of users implicitly.
    Higher factor values indicate less risk of triggering non-cooperative user behavior when agents lead the conversation.
    Thus, when this factor is high, the agent intelligently increases the goal weight to accelerate reaching the goal topic.
    Conversely, the agent decreases the goal weight to avoid triggering non-cooperative user behavior.
    
  \end{itemize}

\subsection{Case Study}
\label{exp4}

In this section, a case study was conducted to better demonstrate the superiority between our proactive dialogue policy and other compared dialogue policies.
Figure~\ref{fig:case} shows four dialogue fragments of \modelname{} and three baselines: Pop (GCR), Pop (US) and \modelname{}-degrade, interacting with a user simulator whose tolerance is 1.0, as well as the final results containing the GCR and US scores.
We obtain the following observations: 
\begin{itemize}[leftmargin=5mm]
    \item[i)] Pop (GCR) baseline works to improve the GCR score solely, so it chooses the topic closest to the goal in each turn.
    However, due to the agent ignores the user satisfaction, the user behaves non-cooperatively in almost every turn.
    ach non-cooperative behavior again takes the conversation off-goal again, resulting in the agent's efforts being wasted.

    \item[ii)] Pop (US) baseline works to improve the US score and therefore introduces the user's favorite topics in each turn. High user satisfaction is maintained during the conversation, resulting in few non-cooperative user behaviors. 
    Due to the neglect of the target \uppercase\expandafter{\romannumeral1}, the GCR score is low at the end of the conversation, despite maintaining a high user satisfaction.

    \item[iii)] In \modelname{}-degrade, the goal weight is a fixed value (after training), i.e., $\beta=0.54$. The $\beta$ value greater than 0.5 means that the agent will always be partial to the goal achievement in the conversation. As we can see, in the first turn, when the user satisfaction is significantly low, the agent still chooses one topic that is close to the goal rather than preferred by the user.
    However, this decision triggers the non-cooperative user behavior in the next turn. 
    Obviously, it is difficult to intelligently balance user satisfaction and goal completion using only one parameter.

    \item[iv)] \modelname{} is significantly more flexible and strategic than the three baselines. It is able to choose to favor US when user satisfaction is low~(e.g., $gw_2=0.07$), while favoring GCR at a high US and later stages of the conversation~(e.g., $gw_{15}=0.55$).
    
  \end{itemize}

  %

%% file: body/relatedwork.tex

\section{Related Work}
\label{relatedwork}

Conventional conversational systems are dedicated to assisting users to accomplish goals or achieve users' satisfactions.
For example, task-oriented systems, always as virtual personal assistants, are designed to help users with daily tasks, such as booking accommodation and restaurant~\cite{yan2017building,liu2018dialogue,kepuska2018next} conversational recommendation systems~\cite{kang2019recommendation,EAR} can recommend products to users that they want; and human-like chatbots can interact with users to provide reasonable responses for entertainment~\cite{Bordes2017Learning,Adiwardana2020Towards}.

In recent years, however, there has been growing interest in exploring on another feature of dialogue systems: proactivity.
Proactive dialogue systems aims to lead the conversation to their own goals, which may not perfectly align well with the user's goal or satisfaction.
Proactive dialogue systems have great potential in various scenarios.
For example, persuasion dialogue system attempts to convince the user to change his/her attitude, opinion, or behavior.
\citet{wang-etal-2019-persuasion} collect a large dialogue dataset in which a participant persuaded another participant to make a donation.
\citet{tian2020understanding} build a classifier to predict users' policies for resisting donations.
Negotiation dialogue system interacts with a user strategically to reach an agreement.
\citet{he-etal-2018-decoupling} propose a modular approach based on coarse dialogue acts that decouples policy and generation to make the policy controllable.
\citet{2019Augmenting} model both semantic and policy history to improve both dialogue policy planning and generation performance. However, 
the proactive policy studied in the aforementioned works lack explainability hence helping less on revealing the rationale of proactive dialogues.


To gain better explainability and empower strategic reasoning for proactive dialogue, \citet{2019Proactive} introduce a knowledge graph and formalize the proactive policy as a path reasoning problem over the KG. They contribute a corpus through crowdsourcing where each dialogue history is composed by two workers, one worker acting as a conversation leader while the other acting as the follower.
Under this setting and the corpus, \citet{Hao2020Multi} propose an End-to-End dialogue model based on Memory network to study natural language response generation;
\citet{bai2021learning} focus on the issue of knowledge coherence in proactive dialogues;
\citet{zhu2021proactive} use a retrieval-based approach for knowledge prediction in proactive dialogues.
However, these efforts all follow a corpus-based learning manner.
Fundamentally different from existing efforts, we take one step further to scrutinize proactive dialogue policy with interactive learning in this paper.

%% file: body/conclusion.tex
\section{Conclusion}
\label{conclusion}

In this work, we study the proactive dialogue policy in an interactive manner and call attention to the non-cooperative user behavior during the conversation.
We argue that the interactive proactive dialogue policy learning has two targets: \textit{leading the conversation to the goal quickly} and \textit{maintaining a high user satisfaction}.
To advance the two targets, we propose \modelname{} which employs a learned goal weight to achieve a trade-off between them.
We design user simulators to interact with the agents during training and evaluation.
The experimental results demonstrate that \modelname{} opens the performance gap for interactive proactive dialogue policy learning.

Our work takes the first step to advance the interactive proactive dialogue policy learning, and can serve as a preliminary baseline to benefit further research.
Naturally, there are thus a few loose ends for further investigation, especially with respect to more diverse user behavior and richer user personalities. 
We will also try to enhance the goal weight by considering more influencing factors.
Lastly, we will deploy \modelname{} to online applications that interact with real users to gain more insights for further improvements.